\documentclass[letterpaper, 10 pt, conference]{ieeeconf}  

\IEEEoverridecommandlockouts                              
\usepackage{physics}
\usepackage{graphicx}

\title{\LARGE \bf
Simulating 0+1 Dimensional Quantum Gravity on Quantum Computers: Mini-Superspace Quantum Cosmology and the World Line Approach in Quantum Field Theory
}

\author{Charles D. Kocher$^{1}$ and Michael McGuigan$^{2}$
\thanks{$^{1}$C. D. Kocher is with the Department of Physics, Brown University, Providence, RI 02912, USA.
        {\tt\small charles\_kocher at brown.edu}}%
\thanks{$^{2}$M. McGuigan is with the Computational Science Initiative, Brookhaven National Laboratory, Upton, NY 11973, USA.
        {\tt\small mcguigan at bnl.gov}}%
}

\begin{document}

\maketitle
\thispagestyle{empty}
\pagestyle{empty}

\begin{abstract}

Quantum computers are promising candidates to radically expand the domain of computational science by their increased computing power and more effective algorithms. In particular, quantum computing could have a tremendous impact on the field of quantum cosmology. The goal of quantum cosmology is to describe the evolution of the universe through the Wheeler-DeWitt equation or path integral methods without having to first formulate a full theory of quantum gravity. The quantum computer provides an advantage in this endeavor because it can perform path integrals directly in Lorentzian space and does not require constructing contour integrations in Euclidean gravity. Also, quantum computers can provide advantages in systems with fermions which are difficult to simulate on classical computers.  In this study, we first employed classical computational methods to analyze a Friedmann-Robertson-Walker mini-superspace with a scalar field and visualize the calculated wavefunction of the universe for a variety of different values of the curvature of the universe and the cosmological constant. We then used IBM's Quantum Information Science Kit Python library and the variational quantum eigensolver algorithm to study the same systems on a quantum computer. The framework was extended to the world line approach to quantum field theory.

\end{abstract}

\section{INTRODUCTION}

In many branches of science, the most interesting and realistic models of physical systems are ones that are not analytically solvable: the mathematical complexity of the problem is too great for an exact solution to be obtained. This reality has led to a need for ever-larger supercomputers in order to increase both the accuracy of current predictions and the number of physical systems that can be studied computationally. Despite the enormity of resources devoted to this problem, building supercomputers powerful enough to simulate physically important systems is still an open problem \cite{vogele2018PhRvL.120z8104V,goh2017arXiv170104503G,davies2005hep.lat...9046D}.

Quantum computing is an emerging technology that has the potential to radically shift this paradigm. A relatively small fully-connected, fault-tolerant quantum computer could perform the same computations as modern supercomputers \cite{biamonte2018arXiv180800460B}. Moreover, because they are quantum systems, quantum computers can exploit the phenomena of entanglement and superposition to run novel algorithms. While full quantum supremacy may be many years away, current quantum computers can perform quantum chemistry calculations on small molecules \cite{kandala2017hardware} and quantum electrodynamics calculations using the Schwinger model \cite{klco2018arXiv180303326K}, among others \cite{cervera2018arXiv180707112C,kapil2018arXiv180700521K,zhukov2018arXiv180710149Z,manabputra2018arXiv180610229M}.

Many private companies have made efforts to become leaders in quantum technology. In particular, IBM has engineered many quantum computers up to 50 qubits in size and has made some of them publicly accessible through their Quantum Information Science Kit (QISKit) Python library \cite{santos2016arXiv161006980S,cross2017arXiv170703429C}. This software includes implementations of many quantum algorithms and simulators for local testing; thus, the application of quantum computational methods to a variety of problems in the sciences is now a reality. In this paper, we present results obtained using QISKit to study quantum cosmological models of the universe. Using classical computational methods to study quantum cosmology has its difficulties, from the restriction to imaginary time to ambiguities concerning integration contours \cite{brown1990lorentzian,wiltshire1996introduction,feldbrugge2017lorentzian,dorronsoro2017arXiv170505340D,feldbrugge2017arXiv170805104F,dorronsoro2018arXiv180401102D,feldbrugge2018arXiv180501609F,tucci2018arXiv180607134D}. Quantum computational methods could ultimately circumvent these issues, providing the motivation for increased efforts to begin the process of simulating quantum cosmology on quantum computers. We begin that process here by analyzing the Wheeler-DeWitt canonical quantization of Friedmann-Robertson-Walker mini-superspaces. Our results indicate that the future of studying quantum cosmology on quantum computers is bright and that new quantum algorithms will allow for even more information to be extracted from these simulations.

\section{QUANTUM COMPUTING}

\begin{figure}[t]
\includegraphics[width=\columnwidth]{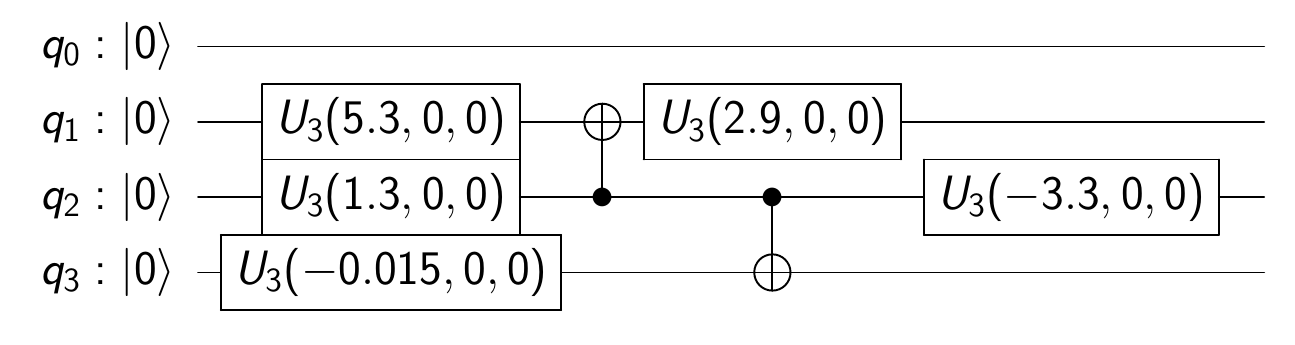}
\caption{\label{fig:dw_opt_circ} The quantum circuit used to generate the wavefunction in Figure~\ref{fig:dw_wf}. The $u3$ rotation gates are represented by boxes on the line corresponding to the qubit they act on, while the CNOT gates are represented by two circles, with the dark circle on the control qubit and the circumscribed plus on the target qubit.}
\end{figure}

Quantum computing exploits the physics of many-particle two-level (qubit) systems in order to perform numerical calculations. In a typical quantum computing calculation, $n$ qubits start in the state $\ket{00...0}$ and are subsequently acted on by quantum gates, which produce a state that is a superposition of the $2^n$ possible tensor product basis states of the qubits. In the Hilbert space of the qubit system, these quantum gates are represented by unitary operators. Examples of gates are the $u3$ operator, which rotates a single qubit in its individual Hilbert space, and the CNOT operator, which acts on two qubits by flipping the state of the target qubit if the control qubit is in the $\ket{1}$ state. The full experimental setup of gates and qubits is known as a circuit; an example of a circuit is shown in Figure~\ref{fig:dw_opt_circ}. After the operations of the gates in the circuit have been performed, the quantum state of the qubit system is measured, returning a string of $n$ ones and zeros. After many preparations and measurements of the system, known as shots, the quantum state produced by the circuit can be determined from the relative frequency of the final measured states.

Many algorithms for quantum computers have been developed to calculate various quantities; one algorithm is the variational quantum eigensolver (VQE) algorithm \cite{kandala2017hardware,peruzzo2014variational}. The VQE algorithm exploits the variational method of quantum mechanics to provide an upper bound for the ground state energy of a Hamiltonian. The Hamiltonian must first be written as a finite $2^n-$by$-2^n$ matrix using the raising and lowering operators of the truncated harmonic oscillator Hamiltonian. The quantum computer is then used to find the expectation value of this matrix given some trial state, which is created using a circuit with tunable parameters. The expectation value is minimized with respect to these parameters using a classical optimization method; for this study, the only optimization method used was the simultaneous perturbation stochastic approximation (SPSA) \cite{spall119632}. After the VQE algorithm completes, an optimized trial circuit and an upper bound for the ground state energy are returned.

\begin{figure}[t]
\includegraphics[width=\columnwidth]{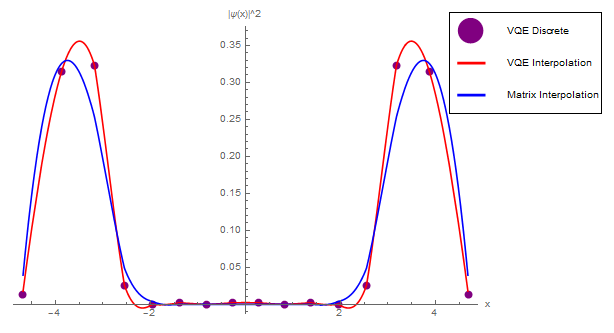}
\caption{\label{fig:dw_wf} Comparison of the predicted probability distributions obtained from the VQE (purple points and red curve) and the matrix method (blue curve) for the double-well matrix Hamiltonian.}
\end{figure}

The VQE algorithm was first used to study three simple quantum mechanical systems in one dimension that are an integral part of some quantum cosmological models of the universe: these systems were single particles in the harmonic oscillator, the anharmonic oscillator, and the double-well potentials. The corresponding Hamiltonians used were 
    
\begin{align}
H_1 &= \frac{1}{2}p^2 + \frac{1}{2}x^2 \; \label{eq:qho} \\
H_2 &= \frac{1}{2}p^2 + \frac{1}{2}x^2 + \frac{0.275}{4}x^4 \; \label{eq:qaho} \\
H_3 &= \frac{1}{2}p^2 - x^2 + \frac{0.15}{4}x^4 \; \label{eq:qdw} 
\end{align}

\noindent where $\hbar$ was set to unity.  The VQE implementation used was included in IBM's QISKit Python package and was run on QISKit’s QASM quantum computer simulator. The VQE-calculated probability distribution for the ground state of the double-well potential is shown in Figure~\ref{fig:dw_wf}, along with its corresponding classical computational solution. The purple points form the discrete probability distribution obtained from the VQE; the red curve is the interpolation of this discrete function into a continuous one. The blue curve is the interpolation of the discrete probability distribution obtained from the classical matrix method. In addition to the close agreement seen with each wavefunction, the VQE upper bounds on the ground state energies $E_0$ were close to the ground state in each case: the full results are listed in Table~\ref{tab:results}. 

\begin{table}

\begin{centering}
\begin{tabular}{|lcr|}
\hline
Potential&Exact $E_0$&VQE Upper Bound\\
\hline
Harmonic Oscillator & 0.5 & $0.50001 \pm 0.00003$\\
Anharmonic Oscillator & 0.543116 & $0.55 \pm 0.09$\\
Double-Well & -5.68592 & $-5.55 \pm 0.05$\\
\hline
\end{tabular}
\caption{\label{tab:results} Comparison of the exact ground state energies of the systems studied to the VQE upper bounds, in units of $\hbar$. All upper bounds were within 3\% of the actual ground state energy.}
\end{centering}
\end{table}

\section{QUANTUM COSMOLOGY}

Quantum cosmology is a method of studying the dynamics of the universe without first formulating a full theory of quantum gravity. When constructing a quantum cosmological universe, the Einstein-Hilbert action of general relativity is quantized using either path integral methods \cite{brown1990lorentzian,feldbrugge2017lorentzian,garay1991path} or canonical quantization and the Wheeler-DeWitt equation \cite{dewittPhysRev.160.1113,hawking1984quantum}. Only the Wheeler-DeWitt canonical quantization was used in this work. 

The Wheeler-DeWitt equation is a constraint on the Hamiltonian of a universe: $H\psi = 0$, where $H$ is the Hamiltonian derived from the Einstein-Hilbert action, and $\psi$, a function over the space of possible 3-geometries and matter field configurations of the universe, is the wavefunction of the universe. In order to render this problem over an infinite-dimensional parameter space tractable, the possible 3-geometries of the universe were restricted to those which have only a finite number of variables because of the high degree of symmetry they possess; this restricted parameter space is known as mini-superspace \cite{vilenkin1993quantum,halliwell1991introductory}. The simplest mini-superspace is the Friedmann-Robertson-Walker (FRW) mini-superspace, which results from the metric 

\begin{equation}
ds^2 = -N^2(t) dt^2 + a^2(t) d\Omega_3^2 \; , \label{eq:frwmetric}
\end{equation}

where $a(t)$ is the scale factor, $d\Omega_3^2$ is the 3-metric for a spatially homogeneous and isotropic universe, and $N(t)$ is the lapse function, which fixes the choice of gauge \cite{pedram2009conformally}. A scalar field or other matter component can be added to the action resulting from the spacetime geometry to model universes with various different characteristics. Three different universes were studied, each with different scalar fields and geometric parameters as described in the following sections. 

\subsection{Closed $\Lambda = 0$ Universe with a Conformally Coupled Free Scalar Field}

For a conformally coupled free scalar, the action is \cite{pedram2009conformally,vilenkin2018arXiv180802032V}

\begin{equation}
S = \int d^4x \sqrt{-g}\left[\frac{R}{16\pi G} - \frac{1}{2}\nabla_\mu \phi \nabla^\mu \phi -\frac{1}{12}R \phi^2 \right] \; , \label{eq:ehaction1}
\end{equation}

where $g$ is the determinant of the FRW metric and $R$ is the curvature  scalar.  Using the variable $\chi = \frac{a \ell_p \phi}{\sqrt{2}}$ where  $\ell_p$ is the Planck length, and choosing the $N(t) = a(t)$ gauge, the Hamiltonian for the system is

\begin{equation}
H = -\frac{p_a^2}{4} + \frac{p_\chi^2}{4} - a^2 + \chi^2 \; , \label{eq:qcho}
\end{equation}

\noindent where $p_a = -2\dot{a}$ and $p_\chi = 2\dot{\chi}$ (in the $N = a$ gauge) are the conjugate momenta to $a$ and $\chi$. The form of this Hamiltonian is essentially that of two copies of $H_1$ (Equation \ref{eq:qho}) that have opposite signs. The result of using the VQE to verify the Wheeler-DeWitt constraint for this universe yielded $E_0 = 0.0000 \pm 0.0003$.

\subsection{Closed $\Lambda < 0$ Universe with a Conformally Coupled $\phi^4$ Scalar Field}

The Hamiltonian for the second universe is

\begin{equation}
H = -\frac{p_a^2}{4} + \frac{p_\chi^2}{4} - a^2 + \chi^2 - |\Lambda|a^4 + c\chi^4 \; . \label{eq:qcaho}
\end{equation}

\noindent This Hamiltonian essentially contains two copies of $H_2$ (Equation \ref{eq:qaho}) with opposite signs. The result of using the VQE to verify the constraint for this universe with $|\Lambda| = c = 0.275/4$ yielded $E_0 = 0.04 \pm 0.05$.

\subsection{Open $\Lambda < 0$ Universe with a Conformally Coupled $\phi^4$ Scalar Field}

The Hamiltonian of the final universe is 

\begin{equation}
H = -\frac{p_a^2}{4} + \frac{p_\chi^2}{4} + a^2 - \chi^2 - |\Lambda|a^4 + c\chi^4 \; . \label{eq:qcdw}
\end{equation}

\begin{figure}[t]
\includegraphics[width=\columnwidth]{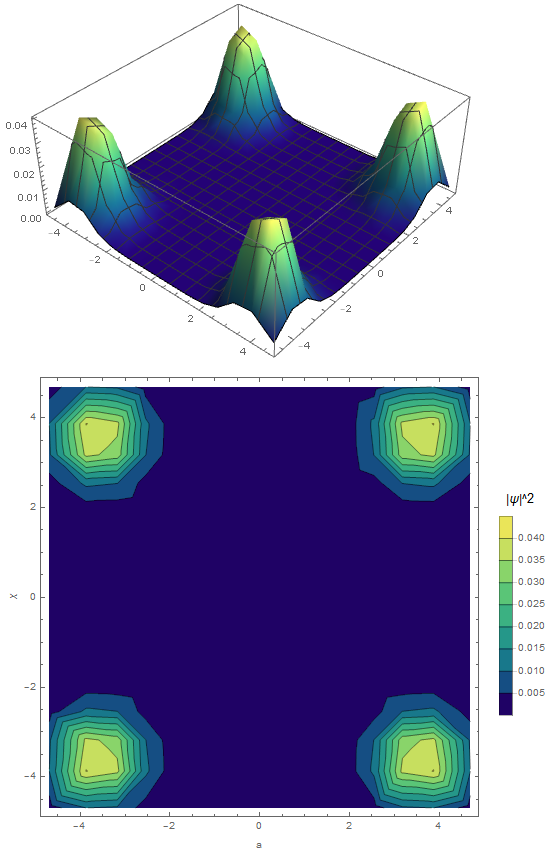}
\caption{\label{fig:qcdw_wf_00} The square of the discrete $\ket{0}\otimes\ket{0}$ wavefunction obtained from the classical matrix method for the Hamiltonian in Equation~\ref{eq:qcdw}.}
\end{figure}

\noindent This Hamiltonian essentially contains two copies of $H_3$ (Equation \ref{eq:qdw}) that have opposite signs. The absolute square of the wavefunction of the universe for a possible zero eigenstate is shown in Figure~\ref{fig:qcdw_wf_00}. The result of using the VQE to verify the Wheeler-DeWitt constraint for this universe with $|\Lambda| = c = 0.15/4$ yielded $E_0 = 0.02 \pm 0.01$.

\section{WORLD LINE QUANTUM FIELD THEORY}

The world line approach to quantum field theory is a string-theory-inspired way to calculate Feynman diagrams. This method utilizes the fact that world sheet Feynman diagrams in string theory can pass to many different QFT Feynman diagrams in the limit that the string tension goes to infinity \cite{schubert1996hep.th...10108S}. The structure of some world line QFT models that are used to study strong interactions is similar to that of quantum cosmology. For example, the Hamiltonian \cite{mueller2017world}

\begin{equation}
H = \frac{\epsilon}{2}(P^2 + m^2 + i \psi^\mu F_{\mu\nu}\psi^\nu) + \frac{i}{2}(P_\mu \psi^\mu + m \psi_5)\chi \; , \label{eq:muellerHWLQFT}
\end{equation}

\noindent where $\epsilon$ and $\chi$ are constraint fields, or the Lagrangian \cite{pisarski1997hep.ph...10370P}

\begin{equation}
L = -i p \cdot \dot{x} - \lambda^\dagger \dot{\lambda} + \frac{1}{2}(p - 2g\Tr(QA))^2 \; , \label{eq:pisarskiLWLQFT}
\end{equation}

\noindent where $\lambda(t)$ are world line fermions in the fundamental representation of $SU(N)_C$ and $Q$ is a nonabelian charge, could be simulated on a quantum computer to determine particle propagation in the field of a gluon. That analysis will be left to a future work.

\section{CONCLUSIONS}

Quantum computing will almost certainly play a large role in the future of scientific computing. In this work, it was shown that the VQE algorithm is capable of producing tight bounds on the ground state energy of quantum systems: all of the bounds for simple quantum mechanical systems obtained above were within 3\% of the exact ground state energies. In the context of quantum cosmology, the VQE was shown to be capable of probing the zero eigenspace of the Hamiltonian to yield wavefunctions consistent with the Wheeler-DeWitt equation. These promising results should motivate further applications of quantum computing to quantum cosmology and structurally related fields, like world line quantum field theory. In the (hopefully near) future, the application of new quantum algorithms, such as QISKit's implementations of quantum phase estimation or dynamics, to quantum cosmological or other systems could give quantum computers a distinct advantage over classical computers in performing computational physics. 




\section*{APPENDIX}

\subsection{\label{sec:charVQE}Characterization of QISKit ACQUA's VQE}

\begin{figure}[t]
\includegraphics[width=\columnwidth]{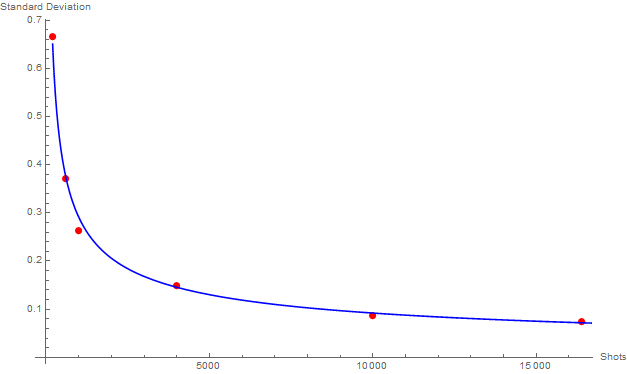}
\caption{\label{fig:shotsanalysis} Plot of the dependence of the standard deviation of the VQE prediction for the anharmonic oscillator Hamiltonian against the number of shots for a fixed circuit simulated on the QASM emulator. The red dots are measured values; the blue curve is the function $\frac{9.21031}{\sqrt{x}}$, which was obtained from fitting the curve $\frac{A}{\sqrt{x}}$ to the data.}
\end{figure}

A short analysis of the behavior of the VQE as a function of the number of shots was performed in order to ensure that the optimization procedure would operate with as little noise as possible. Figure~\ref{fig:shotsanalysis} shows the result of this analysis. The red points are the standard deviations calculated from 100 measurements of the expectation value of the anharmonic oscillator Hamiltonian for a fixed circuit performed on the QASM emulator with the number of shots indicated on the horizontal axis; the blue curve is the function $\frac{9.21031}{\sqrt{x}}$, which was obtained from fitting the curve $\frac{A}{\sqrt{x}}$ to the data. In order to balance available computing power with accuracy, it was determined that 8192 shots would give the best results for calculations.

\subsection{\label{sec:repro}Analysis Code}

Most of the code used to obtain the results in this work is available via the following Github repository: https://github.com/CharlesKocher/simulating-mss-qc-on-quantum-computers. 

\section*{ACKNOWLEDGMENTS}

This project was supported in part by the U.S. Department of Energy, Office of Science, Office of Workforce Development for Teachers and Scientists (WDTS) under the Science Undergraduate Laboratory Internships Program (SULI). We acknowledge use of the IBM Q for this work. The views expressed are those of the authors and do not reflect the official policy or position of IBM or the IBM Q team. The authors would also like to thank Peng Liu for enlightening discussion regarding the VQE algorithm and its QISKit implementation. Michael McGuigan is supported from DOE HEP Office of Science DE-SC0019139: Foundations of Quantum Computing for Gauge Theories and Quantum Gravity.

\nocite{*}
\bibliographystyle{ieeetr}
\bibliography{references}

\end{document}